# Electronic properties and stability of three new kinds of single-atom-thick SIC graphyne sheets

Xiao Yan[a], Zihua Xin[a], M. Yu[b], Jiaojiao Zhang[a], Mihai Deng[a], Lijun Tian[a]

[a] Department of Physics, Shanghai University, Shanghai, 200444, China

[b] Department of Physics and Astronomy, University of Louisville, Louisville, KY 40292, USA

**Abstract**

Three new single atom layered silicon-carbon stable systems have been found by using of SCED-LCAO and DFT methods. An important position, named bone position, is proposed in these structures. For SiC and $Si_1C_9$ system, the bone positions are partially occupied by Si atoms, the plane structure is kept and electronic gap is opened with 0.955 eV and 0.689 eV respectively. For $Si_2C_8$ system, the bone positions are fully occupied by Si atoms. It shows a buckled structure with a buckling of 0.05 Å and a Dirac cone at M point. Moreover, the *sp* hybridization between Si and C atoms in SiC system is found and the co-existence of *sp*, *sp*$^2$ and *sp*$^3$ hybridization is also found in $Si_2C_8$ system. The thermal stability for these three systems is certified.

## 1. Introduction

In recent years, along with the extensive researches on the graphene, the other 2D carbon allotropes have also become the focus of hot topics. Among them, the graphyne is the most interesting one,[1-4] since it is considered to be the most stable compound among artificial, unnatural carbon allotrope.[5] Not only has it abundant carbon bonds, big conjugate structure, wide spacing, but it also has excellent chemical stability and semiconductor performance. It makes the graphyne promised material and having potential applications in electronics, energy storing, and optoelectronics owing to its special electronic structure.[6-9] According to the DFT (density functional theory) simulations, Malko and Neiss[10] showed that the graphyne has a better conductivity than the graphene. The other research on the electronic structure of the multilayer graphyne ribbon also show that the nanoribbons are semiconductors with suitable band gaps similar to silicon and the gaps decrease as widths of nanoribibbons increase.[11]

Graphdiyne, a member of the graphyne family, has been realized experimentally by Li et al.[6] They have successfully synthesized the large area of graphdiyne films (with 3.61 cm$^2$) on the surface of copper via a cross-coupling reaction using hexaethynylbenzene.[6] The SEM and TEM results show that this film can grow continuously on large surface area of copper and has a good flexibility. This work also shows that the graphdiyne has a plane structure with a few defects. The conductivity of this film is 2.516×10$^{-4}$ S/m. Jiao[12] et al. calculated the capacity of graphdiyne surface adsorption for H$_2$ and CH$_4$, the results indicate that the adsorption energy has a maximum value when the gas molecule is in the center of the cell of graphdiyne. Zheng[13] et al. studied the properties and electronic structure of the multilayer graphdiyne using first principle method. They gave the stable multilayer graphdiyne structure and the dependence of gap on the intensity of electric field.

The successful synthesis of the graphdiyne also stimulated the study on the 2D derivatives of the graphyne. A series of intersting properties such as the superconductivity, electricity and mechanics are predicted.[14, 15] Enyashin et al.[14] simulated the stability and electronic structure of the graphyne doped with Fluorine, and found that its properties are highly dependent on stoichiometry of C and F elements. Ongun et al.[15] forecasted a new 2D BN sheet which is similar to the structure of the α-graphyne and simulated the heating process and hydrogenation process for obtaining the electronic structure and magnetic properties. They demonstrated that α-BNyne nanostructures can form stable and durable 2D extended structures with interesting chemical and physical properties.

SiC based materials, one of the interesting semiconducting materials, are widely used in aviation, aerospace, automobile, machinery, electron engineering and chemical engineering industry owing to its excellent performance of thermal conductivity，abrasion, and corrosion resistance. It is also found that SiC nanostructures have more outstanding properties,[16-19] which attracted tremendous attentions. For instance, Sun[20] made a deep study on the synthesizing SiC nanowires (structure with $sp^3$ type of the hybridization) and SiC nanotubes (structure with $sp^2$ type of the hybridization) of SiC, and Yu et al.[21] studied the stability of 2D SiC honeycomb structure. They found that the $sp^2$ hybridization could exist between Si and C atoms in the SiC graphitic-like structures. These previous studies indicate that the $sp^2$ and $sp^3$ hybridization can be formed between Si atom and C atoms.

Now it is very interesting to know what will happen when the Si atoms are introduced into graphyne system. Will such single layer sheet be stable? Could the $sp$ hybridization form between Si and C atoms? Will such systems possess energy gap and show the semiconductor nature? To answer these questions, we have systematically studied Si$_m$C$_n$ single layered systems by substituting C atoms with Si atoms in the framework of the graphyne structure. We have found three stable carbon-silicon graphyne systems, namely, SiC graphyne, $Si_1C_9$ graphyne, and $Si_2C_8$ graphyne using a quantum mechanics based molecule dynamic scheme (referred as SCED-LCAO MD [22,23]). In the following section, we will first briefly describe our simulation scheme . In section 3, we will discuss the structural stability, the electronic band structure, and chemical bonding nature of such conjugate system in terms of their composition and hybridization. In the meantime, the thermal stability is also studied. The conclusion will be finally given in section 4.

**2. Molecular dynamics simulation schemes**

To realize the goal illustrated above, more than 120 initial structures need to be performed in the MD simulations including finite temperatures simulations. It will be a daunting task to survey these many initial structures with *ab-initio* simulations based on the density functional theory (DFT). Therefore, we will employ an efficient molecular dynamics scheme based on a semi-empirical Hamiltonian developed by Louisville group.[22,23] The semi-empirical Hamiltonian, referred as SCED-LCAO, is developed in the framework of a linear combination of atomic orbitals (LCAO). As can be seen from the Eq.1a and 1b in Ref. 23 that the SCED-LCAO Hamiltonian includes environment-dependent (ED) interactions and the approach allows a self-consistent (SC) treatment of charge redistribution. The electron-electron interactions and the electron-ion interactions are described by parameterized functions. The total energy (see Eq.11 in Ref. 23) includes the band structure energy, the correction term from the double-counting of electrons, and ion-ion repulsions, where the band structure energy is obtained by solving a generalized eigenvalue equation. The force acting on each atom (see the Eq.12 in Ref. 23) can be carried out by the Hellmann-Feynman theory.[24,25] The most outstanding feature of SCED-LCAO approach is that it can be employed to optimize structures of large complex systems that are beyond the scope of the *ab*-initio calculations. The comparison on the efficiency between *ab*-initio calculations and SCED-LCAO method can be referred to the Ref. 23. SCED-LCAO molecular dynamics method has been successfully used to study silicon-based systems,[21] large-scale fullerenes [26] and some silicon carbide nanostructures.[27, 28] The Refs. 22, 23 have already elaborated SCED-LCAO method and there is also a brief introduction about the method in the references and appendix of the Refs. 27, 28. In this work we first checked the graphyne using the SCED-LCAO method. The optimized C≡C and C-C bond lengths are 1.278 and 1.436 Å, respectively. The results are in consistent with the DFT results with the full potential linear combination of atomic orbitals method [29] (i.e., 1.220 and 1.396 Å for C≡C and C—C bond lengths, respectively). We then employed the SCED-LCAO method to study the $Si_mC_n$ graphyne structures in the present work. The time step in this molecular dynamics simulation is set to be 1.5 fs. The energy was converged to within $10^{-5}$ eV, and the force criteria for a fully relaxation process was set to be less than $10^{-4}$ eV/Å. To validity our simulation results, the DFT based calculation using VASP software [30-32] is also performed. We used PAW for the pseudopotential and LDA for the electronic exchange correlation. The criteria for the energy convergency is $10^{-5}$ eV with the plane-wave cutoff energy of 540 eV and the criteria for the force in the relaxation process is $10^{-4}$ eV/Å. The Monkhorst-Pack scheme is used to sample the Brillouin zone, and a mesh of $7 \times 7 \times 1$ k-point sampling is used for our simulation. The distance between two layers in the supercell is 15 Å to avoid the interlayer interactions.

## 3. Results and discussions

### 3.1 Structural stability

We have performed a systematic simulation on modeling carbon-silicon graphyne with different stoichiometry including SiC graphyne, $Si_1C_9$ graphyne, $Si_2C_8$ graphyne, $Si_8C_2$ graphyne and $Si_9C_1$ graphyne, and found three stable structures. The schematic structures of the obtained stable $Si_mC_n$ graphyne are shown in Fig. 1(a), (b), and (c), respectively. The grey balls represent the carbon atoms and the yellow ones are Si atoms. In the study of the stability of these structures, we choose a $2\times2$ supercell with the periodic condition on XY plane, and set a big vacuum along the Z direction to make sure that the layers are independent each other. Two types of the initial configurations are constructed for each of these $Si_mC_n$ graphyne systems, one is a flat sheet, and the other is a bucked sheet with the buckling between 0.01-0.5 Å. Then these initial structures are fully relaxed using the SCED-LCAO method mentioned above.

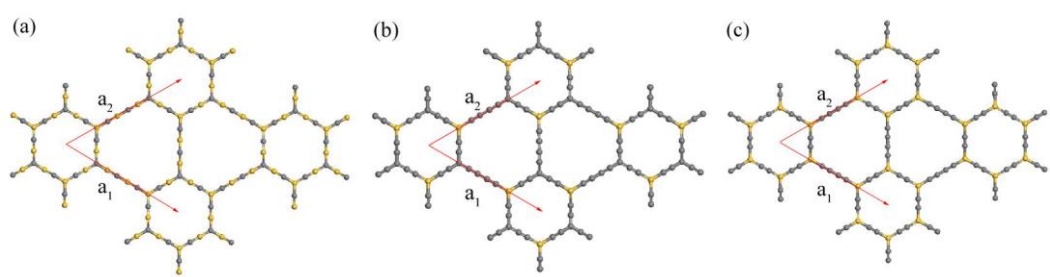

Fig. 1 Structures of (a) SiC graphyne, (b) $Si_1C_9$ graphyne, and (c) $Si_2C_8$ graphyne. The vectors ($a_1$ and $a_2$) in each structure represent the lattice vectors. The grey balls represent the carbon atoms and the yellow ones are Si atoms.

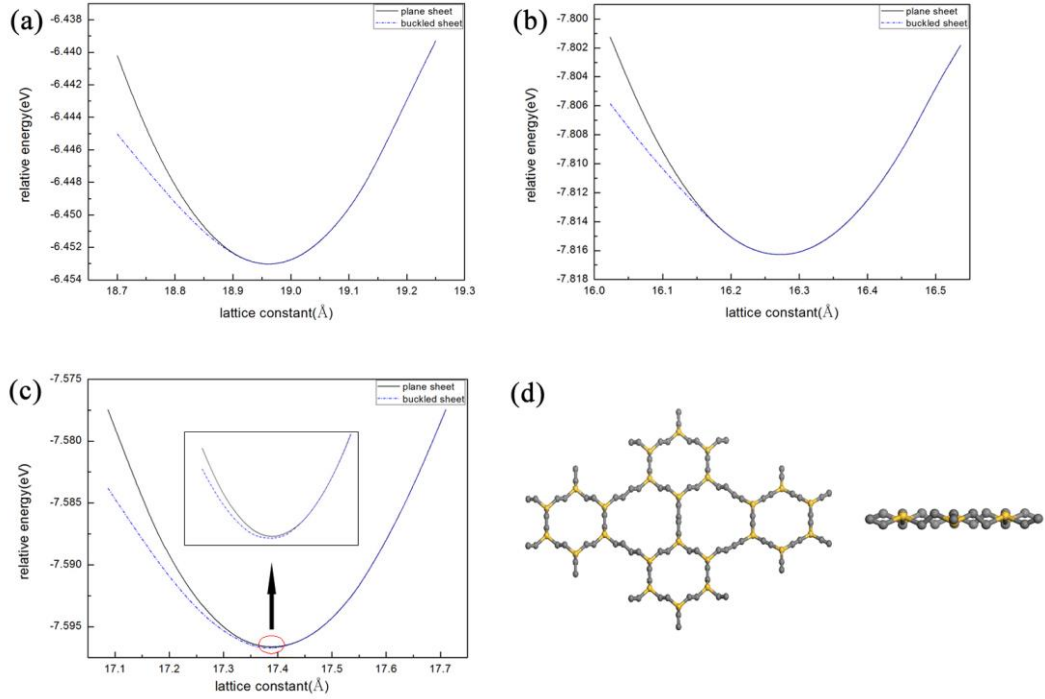

Fig. 2 The relative energy per atom (see the definition in the text) versus lattice constant of (a) SiC graphyne, (b) $Si_1C_9$ graphyne, and (c) $Si_2C_8$ graphyne, respectively. The solid curves are the results for flat layer and the dashed curve for the buckled layer. The inset in (c) shows the energy curves of $Si_2C_8$ around the minimum indicating that the buckled layer is more stable than the flat layer. The top and side views of the stabilized buckled $Si_2C_8$ graphyne is shown in (d).

The obtained relative energy $E_{re}$ per atom versus lattice constant $a$ for the relaxed SiC graphyne, $Si_1C_9$ graphyne, and $Si_2C_8$ graphyne are given in the Fig.2(a)-(c) (the solid curves for the initial flat sheets and the dashed curves for the initial buckled sheets, respectively). The relative energy is defined as $E_{re}=E_{total}-N_{Si}*E_{Si}-N_C*E_C$, where $E_{total}$ is the total energy of the $Si_mC_n$ graphyne system, $N_{Si}$, the number of Si atoms, $E_{Si}$, the energy of single Si atom, $N_C$, the number of C atoms, and $E_C$, the energy of single C atom. What we found is that for SiC graphyne and $Si_1C_9$ graphyne, both the initial flat and bucked sheets will finally stabilized to a flat sheet when $a$ >18.887 Å for SiC graphyne and $a$ >16.163 Å for $Si_1C_9$ graphyne. While, in the case of smaller value of $a$, the buckled sheets are lower in energy than the flat sheets. The optimized lattice constant is 18.956 Å for the SiC graphyne, and 16.279 Å for the $Si_1C_9$ graphyne, respectively. Both the initial flat and buckled sheets are finally stabilized to a flat sheet at the equilibrium and have the same relative energy. The above results indicate that the flat structure is the most stable structure for SiC graphyne and $Si_1C_9$ graphyne. The $Si_2C_8$ graphyne, however, shows a special character of geometrical properties. It can be seen from the inset in Fig.2(c) that the relative energy of $Si_2C_8$ graphyne with the buckled structure is lower than the flat layer around the minimum (17.381 Å), indicating that the most stable structure at the equilibrium is the buckled sheet with the buckling of 0.05 Å. The top and side view of the stabilized buckled $Si_2C_8$ graphyne around the minimum is shown in Fig. 2 (d). The obtained stable flat SiC and $Si_1C_9$ sheets and the buckled $Si_2C_8$ sheet are also confirmed with DFT (LDA-PAW) calculations using VASP software [30-32]. The optimized lattice constants are 18.642 Å and 15.610 Å for flat SiC and $Si_1C_9$ sheets and 16.729 Å for buckled $Si_2C_8$ sheet, respectively. The obtained buckling for $Si_2C_8$ sheet is 0.01 Å.

What we found is that even though the concentration of Si affects the geometry of stable structure, most importantly, the corner position of the hexagon in the structure (referred as the bone position) will play a crucial role. It is found that in the case of $Si_2C_8$ graphyne structure, all the bone position are occupied by Si atoms, so a small buckling appears in this structure. In the case of the SiC graphyne, even though the concentration of Si for SiC graphyne is a little higher than that for $Si_2C_8$ graphyne, its bone position is occupied by C and Si alternatively, and it stabilized to a flat structure.

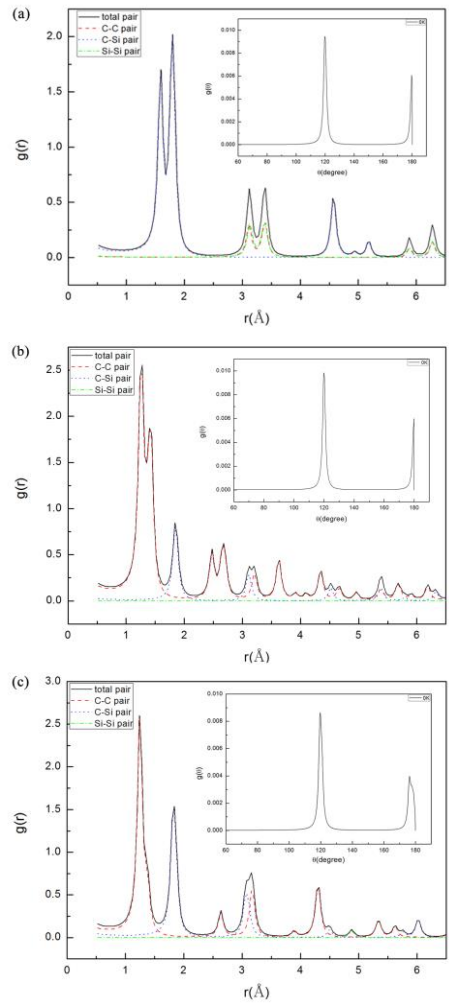

Fig. 3 Total and partial pair distribution functions of (a) SiC graphyne, (b) $Si_1C_9$ graphyne, and (c) $Si_2C_8$ graphyne, at 0 K. The inserts are the total angle distribution functions for each corresponding structure.

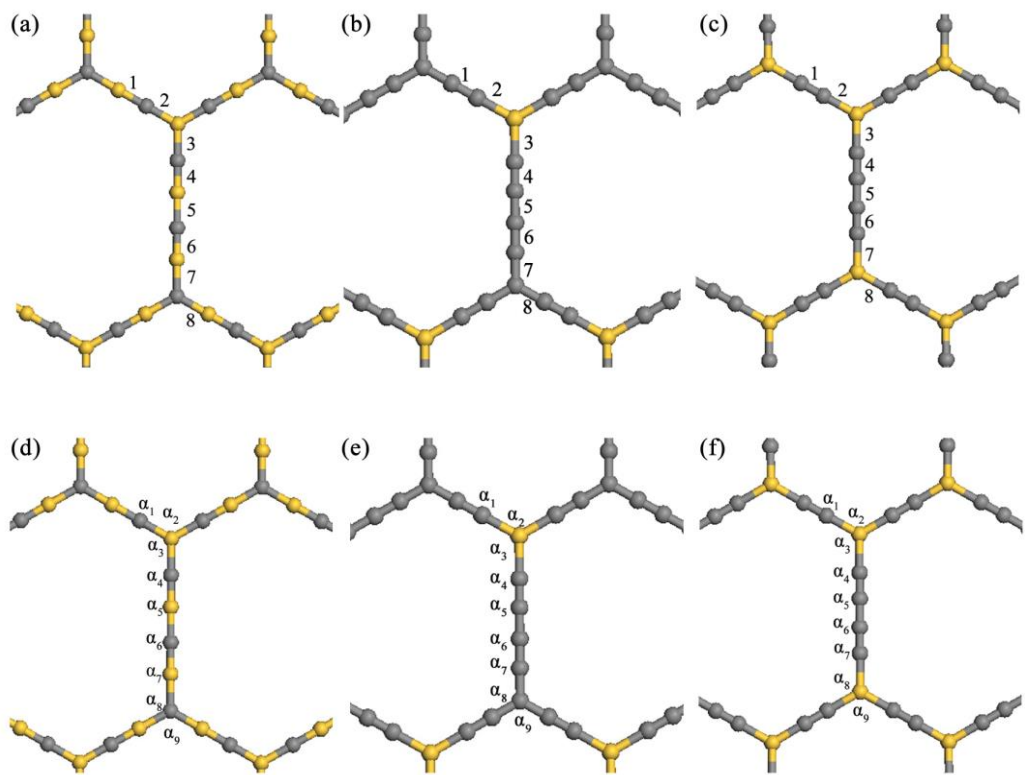

Fig. 4 Si-C and C-C bonding nature of (a) the SiC graphyne, (b) $Si_1C_9$ graphyne, and (c) $Si_2C_8$ graphyne, respectively. The indexes 1-8 denote the locations of various bonds. The angle nature of (d) the SiC graphyne, (e) $Si_1C_9$ graphyne, and (f) $Si_2C_8$ graphyne are indicated by $α_1$-$α_9$, respectively.

To shed light into structural properties, we performed the structural analysis on the obtained stable structures of SiC graphyne, $Si_1C_9$ graphyne and $Si_2C_8$ graphyne, through the pair distribution function, the angle distribution function, and the local structure analysis including the hybridization on bond forming. Fig.3(a)-(c) show the pair-distribution functions for SiC graphyne, $Si_1C_9$ graphyne, and $Si_2C_8$ graphyne respectively. The insets are the angle distributions function for each structure. The first peak in each figure in Fig.3 represents the triple bond in the chain portion of the graphyne structure. The second peak represents the $sp^2$ bonds between the atom at the corner and its nearest neighbor as well as the single $sp$ bonds. The bonds angles, indicated by the peaks in the angle-distribution functions around 120 ° and 180 °, are typical values of $sp$ and $sp^2$ types of the hybridization. In the case of the SiC graphyne, only Si-C bonds exist and the they form both the *sp* and *sp*$^2$ type of bonds (see the 1$^{st}$ and 2$^{nd}$ peaks in Fig. 3 (a)) In the case of the Si$_1$C$_9$ graphyne, there are three types of bonds: the *sp* and *sp*$^2$ types of C-C bonds and the *sp*$^2$ type of Si-C bonds, respectively (see the 1$^{st}$, 2$^{nd}$, and 3$^{rd}$ peaks in Fig. 3(b)).

| Bond index | Bond length (Å) | | | |
| --- | --- | --- | --- | --- |
| | SiC graphyne | Si$_1$C$_9$ graphyne | Si$_2$C$_8$ graphyne | graphyne |
| 1 (*sp*) (triple) | 1.598 | 1.258 | 1.247 | 1.282 |
| 2 (*sp*$^2$) | 1.788 | 1.843 | 1.818 | 1.428 |
| 3 (*sp*$^2$) | 1.821 | 1.874 | 1.864 | 1.453 |
| 4 (*sp*) (triple) | 1.587 | 1.259 | 1.256 | 1.275 |
| 5 (*sp*) (single) | 1.770 | 1.381 | 1.380 | 1.394 |
| 6 (*sp*) (triple) | 1.582 | 1.274 | 1.256 | 1.275 |
| 7 (*sp*$^2$) | 1.831 | 1.438 | 1.864 | 1.453 |
| 8 (*sp*$^2$) | 1.797 | 1.426 | 1.818 | 1.428 |
| Angle index | Angle degree | | | |
| | SiC graphyne | Si$_1$C$_9$ graphyne | Si$_2$C$_8$ graphyne | graphyne |
| $\alpha_1$ | 180 | 180 | 176.1 | 180 |
| $\alpha_2$ | 120.42 | 120.54 | 120.7 | 120.5 |
| $\alpha_3$ | 119.79 | 119.73 | 119.55 | 119.75 |
| $\alpha_4$ | 180 | 180 | 177.66 | 180 |
| $\alpha_5$ | 180 | 180 | 178.7 | 180 |
| $\alpha_6$ | 180 | 180 | 178.7 | 180 |
| $\alpha_7$ | 180 | 180 | 177.66 | 180 |
| $\alpha_8$ | 119.7 | 119.73 | 119.55 | 119.75 |
| $\alpha_9$ | 120.6 | 120.54 | 120.7 | 120.5 |

Table 1 Comparison of various bond lengths and the angles of SiC graphyne (2$^{nd}$ column), Si$_1$C$_9$ graphyne (3$^{rd}$ column), and Si$_2$C$_8$graphyne (4$^{th}$ column) with those of the graphyne (5$^{th}$ column)

While, in the case of the Si$_2$C$_8$ graphyne, the C atoms only form the *sp* type of bonds, and Si and C form *sp*$^2$ type of bonds (see the 1$^{st}$ and 2$^{nd}$ peaks in Fig. 3 (c)). To clearly understand the hybridization properties, we carried out the detail analysis on different types of the bond lengths and bond angles which are presented in Fig.4 and Table 1, respectively. The indexes in Fig. 4 (a)-(c) are denoted for different type of bonds, and the indexes in Fig. 4 (d)-(f) denote the locations of various angles. It is found that in all these three structures, bonds 1, 4, and 6 represent triple bonds and belong to the *sp* type of the hybridization, and the bond 5 is the single bond and also in the sp type of the hybridization, while, bonds 2, 4, 7, and 8 represent *sp*$^2$ type of the hybridization. For comparison, we also analyzed the stable structure of the pure carbon system, referred as 18, 18, 24-graphyne.[33] From Table 1, one can see that the bond angle $\alpha_2$ and $\alpha_3$ (shown in Fig.4 (d)) for SiC graphyne are 120.42 ° and 119.79 °. This nature is also found in the graphyne (see the 5[th] column in the Table 1) indicting that the hybridization of atom at the corner of the ring in the graphyne structure is slightly different from the graphene. This is due to the difference between three neighbor chains at the corner. This asymmetry affects charge distribution in the system, and results in such hybridization. The similar hybridization can be obtained for $Si_2C_8$ and $Si_1C_9$. The calculated bond length of the triple bonds 1,4 and 6 for SiC graphyne are 1.598 Å, 1.587 Å and 1.582 Å, respectively, which are in good agreement with the previous calculation.[34] The bond length indicates that delocalized π-bonds exist in this system.

It is interesting that the bond angles $\alpha_1$, $\alpha_4$ to $\alpha_7$ are exactly 180 ° in the case of the SiC graphyne. The results reveal that the Si and C atoms can form not only the $sp^2$ type of bonding in the SiC graphitic-like structures and $sp^3$ type of bonding in the SiC bulk structures, but also the $sp$ type of bonding in the chain-like structures. That means SiC, like carbon, can have different types of bonding and form various allotropes. Furthermore, it is found that in the case of the slightly buckled $Si_2C_8$ graphyne (with the buckling of 0.05Å), these angles are not 180 °, but are 176.1 °, 177.66 ° and 178.7 °, which can be also seen from its angle distribution (i.e., the peak around 176 ° in the insert of Fig. 3 (c)). The results indicate again that the most stable structure of $Si_2C_8$ graphyne is quasi-2D structure, and the $sp$ hybridization in $Si_2C_8$ graphyne is slightly different from that in SiC graphyne and $Si_1C_9$ graphyne.

**3.2 Electronic structure**

The electronic properties of SiC graphyne, $Si_1C_9$ graphyne, and $Si_2C_8$ graphyne are analyzed from their band structures and electronic partial densities of states (PDOS). Fig.5 (a)-(c) and Fig. 6 (a)-(c) show calculated electronic band structures and PDOS of the optimized SiC graphyne, $Si_1C_9$ graphyne, and $Si_2C_8$ graphyne, using the DFT-based VASP method.[30-32] The local-density approximation (LDA) is adopted for the exchange correlation potential. It has been found that even though the three layered systems have the same graphyne-like structure; their electronic properties are quite different. It can be seen from Fig. 5 (a) that, differ from the graphyne, the SiC graphyne is an indirect band gap (e.g., from K to Γ points) semiconductor material with the band gap of 0.955 eV. The bands are flat and the band widths are narrow, indicating that electrons are more localized in the SiC graphyne system. The PDOS (see in Fig. 6 (a)) clearly show that the *p*-bands including the *x*-, *y*-, and *z*-components mainly dominate the bands at the top of the valence band, while, the *s*-bands dominate the band at the bottom of the conduction band.

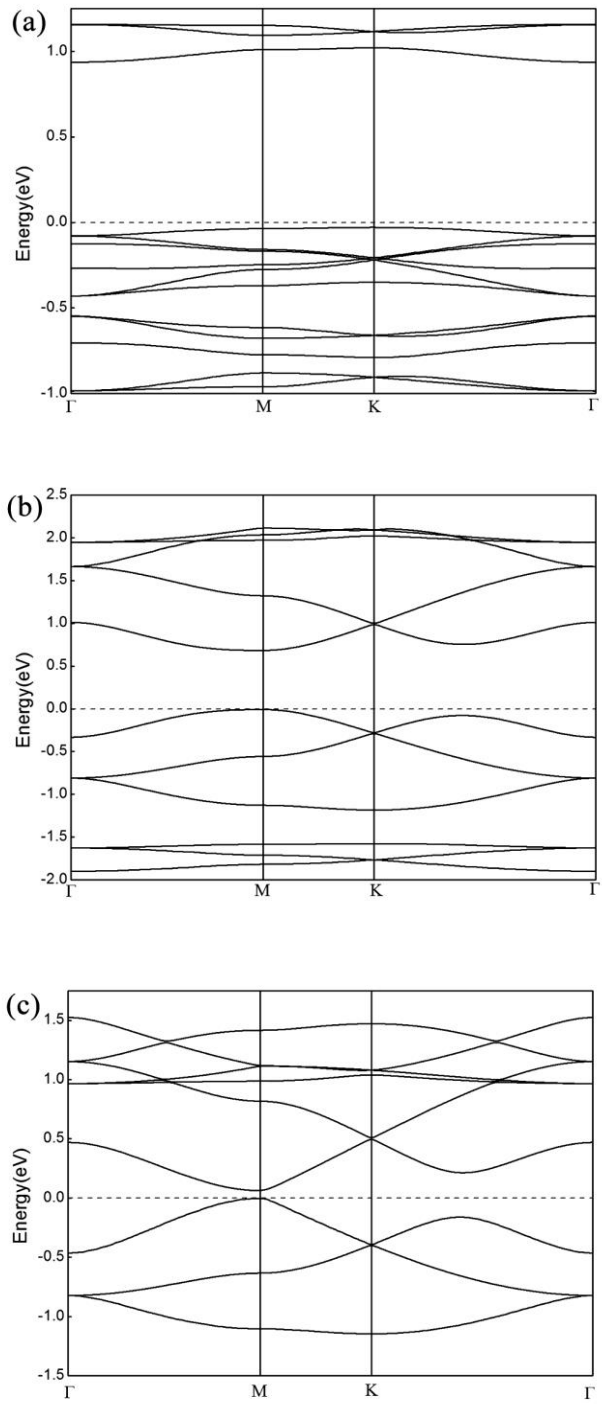

Fig. 5 The band structures of the optimized (a) SiC graphyne, (b) $Si_1C_9$ graphyne, and (c) $Si_2C_8$ graphyne, respectively. The Fermi level is indicated by the dashed lines.

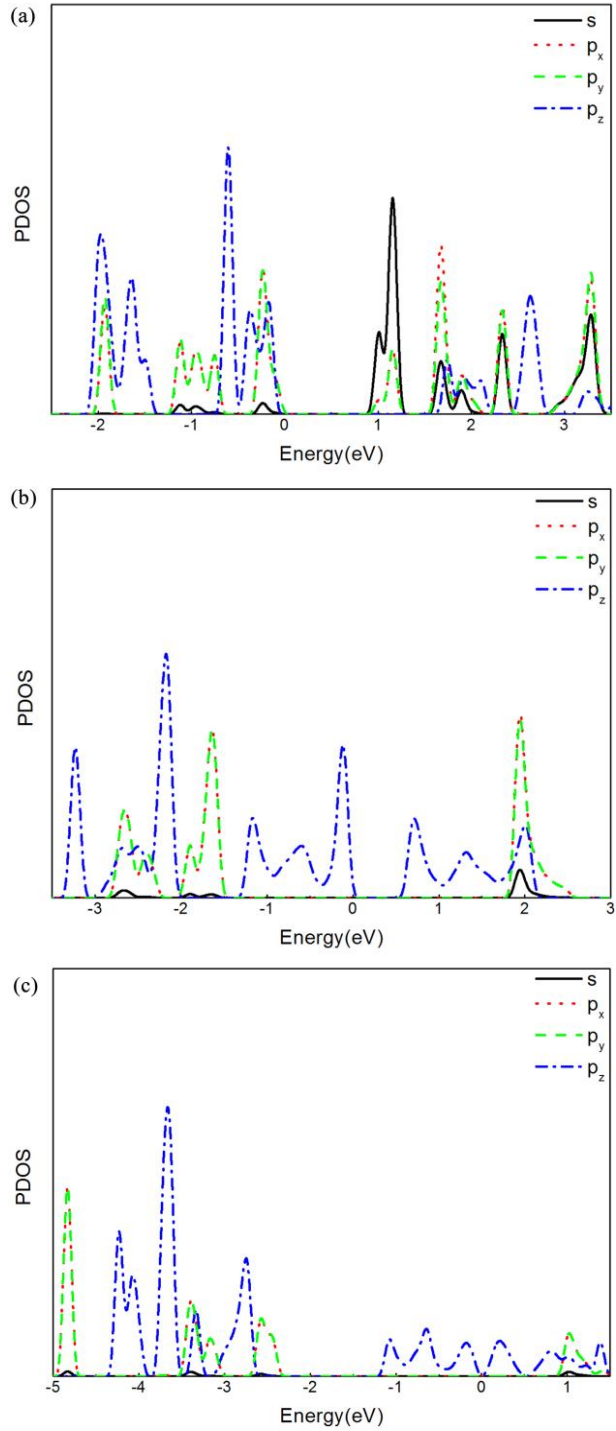

Fig. 6 The partial density of states (PDOS) of the optimized (a) SiC graphyne, (b) $Si_1C_9$ graphyne, and (c) $Si_2C_8$ graphyne, respectively. The Fermi energy is set at zero.

On the other hand, the $Si_1C_9$ graphyne is a direct band gap (e.g., from M to M points as shown in Fig.5 (b)) semiconductor material with the band gap of 0.689 eV. The valence and conduction band shapes near the Fermi level show parabolic nature and are symmetric. The band widths are wider than those in the case of the SiC graphyne, indicating that the electrons in this system are less localized. Compared to the PDOS of the SiC graphyne, it is found that in the case of the $Si_1C_9$ graphyne (see Fig. 6 (b)) only the *z* components of the *p*-bands dominate the bands at the top of the valence bands and the bottom of the conduction bands. In another word, there are no *s*-states and the *p*-states along the *x*- and *y*-directions near the Fermi level.

Quite different from the SiC graphyne and the $Si_1C_9$ graphyne, $Si_2C_8$ graphyne is a gapless material with delocalized nature. In particular, a Dirac cone like band is found. But such Dirac point is found at M point near the Fermi level (see Fig. 5(c)), which is different from the graphene and the α-graphyne at the K point.[10] Such Dirac cone properties can also be seen from the symmetric nature at the Fermi level of the PDOS (Fig. 6 (c)). Furthermore, from the PDOS we can find that, similar to the case of the $Si_1C_9$ graphyne, the states near the Fermi level (e.g., the bands at the top of the valence bands and the bottom of the conduction bands) are mainly from the *z*-components of the *p*-bands. From the above results, we found that the electronic band structure of $Si_mC_n$ based graphyne systems not only depends on the composition (i.e., m/(m+n)) but also strongly depend on the distribution of the Si atoms in the framework of the structures.

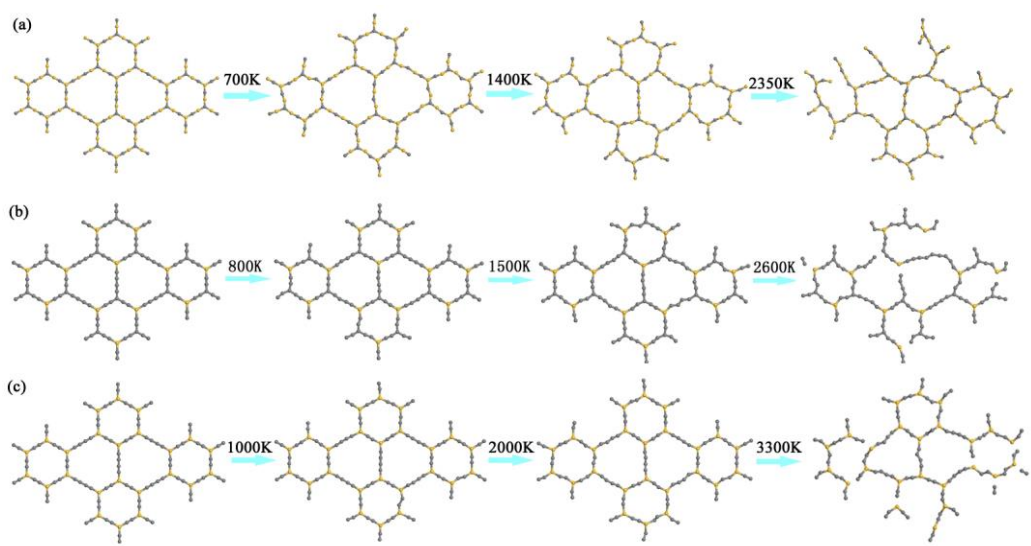

Fig.7 The snapshots of (a) stable flat SiC graphyne, (b) stable flat $Si_1C_9$ graphyne, and (c) stable buckled $Si_2C_8$ graphyne at various stages during the heating processes, respectively.

**3.3 Thermal stability**

In order to study the thermal stability of obtained stable structures, we gradually heat the $Si_mC_n$ graphyne systems up to 3300 K. Fig.7 shows the snapshots of SiC graphyne plane sheet, $Si_1C_9$ graphyne plane sheet and $Si_2C_8$ graphyne buckled sheet at various stages with increasing the temperature, respectively. It is found that these sheets are very stable when the temperature is below 700 K for the SiC graphyne sheet, 800 K for the $Si_1C_9$ graphyne sheet, and 1000 K for the $Si_2C_8$ graphyne buckled sheet, respectively (see the 2$^{nd}$ column structures in Fig.7). Above the these temperatures, these sheets begin to distort (see the 3$^{rd}$ column structures in Fig.7) and then break when the temperature reaches 2350 K for the SiC graphyne sheet, 2600 K for the $Si_1C_9$ graphyne sheet, and 3300 K for the $Si_2C_8$ graphyne buckled sheet (see the last column structures in Fig. 7), respectively. Apparently, the $Si_2C_8$ graphyne buckled sheet is thermally more stable than other two structures.

Furthermore, we found that in the case of the SiC graphyne, a slightly change in bond length and bond angle was found when the temperature is above 700 K but lower than 1400 K, indicating a slight deformation in carbon-silicon honeycomb rings and carbon silicon chain. From our detail analysis in the thermal kinetic process, we found that the thermal vibration of the $sp^2$ Si-C bond in the hexagonal rings is strong when the temperature is about 1400 K and breaks the hexagon symmetry. Such symmetry broken induces the further distortion of the $sp$ bonds (see the structures of SiC graphyne at 1400 K in Fig. 7 (a)). When the temperature increases to 2350 K, the SiC bonds begin to break and lead the plane structure damaged with the maximum value of the buckling to be 1.4 Å (see the structure of the SiC graphyne at 2350 K in Fig. 7 (a)). In the case of the $Si_1C_9$ graphyne, we found that at 800 K, it still keeps its basic planer configuration. Even though the hexagonal rings are slightly deformed at 1500 K (see the structure of the $Si_1C_9$ graphyne at 1500 K in Fig. 7 (b)), no buckling appears until 2200 K. With the temperature increases to 2600 K, fractures occur first at the corners of the rings which are occupied by Si atom, and the cracking at the corner occupied by C atom occurs later (e.g., the structure of the $Si_1C_9$ graphyne at 2600 K in Fig. 7 (b)). Namely, the breaking occurs first from the SiC bonds at the hexagonal rings. In the case of the buckled $Si_2C_8$ structure, we found that it keeps its symmetry till 1000 K (see the structure of the $Si_2C_8$ at 1000 K in Fig. 7 (c)). The C-C bond in the hexagonal rings has a little deformation at 2000 K (see the structure of the $Si_2C_8$ graphyne at 2000 K in Fig. 7 (c)). These deformations gradually lead the $sp^2$ hybridization between Si-C to the $sp^3$ hybridization at high temperature. When the temperature reaches to 3300 K, the structure is damaged with the corrugation reaching about 1.6 Å (e.g., the structure of the $Si_2C_8$ graphyne at 3300 K in Fig. 7 (c)).

The reason why the $Si_2C_8$ structure is thermally more stable than the other two structures is because in the $Si_2C_8$ structure, there six Si atoms at the bone positions. Such high symmetry makes the $Si_2C_8$ structure difficult to be broken in the heating process. SiC graphyne and $Si_1C_9$ graphyne, however, have lower symmetry in terms of the distribution of the Si atoms on the bone position, and the difference of the thermal vibrations between the lighter carbon atoms and heavier Si atoms at the hexagonal rings leads to the structures broken easier. From the detailed study on the hybridization we also found that the bond breaking of the $Si_mC_n$ graphyne systems always happens on the $sp^2$ SiC bonds at the bone position with the increasing of the temperature. Then these breaking lead to the further distortion and breaking on the $sp$ bonds and finally, the systems are totally distorted.

4.**Conclusions**

We have found three stable carbon-silicon graphyne structures. The existence of these novel structures has been systematically studied through the structural relaxation, optimization, electronic structures, as well as the thermal stability. A bone position (i.e., the site at the corner of honeycomb) is proposed in analyzing the structural and bonding nature properties on this structure. From the result, one can see that the more the bone position is occupied by carbon, the more possible to form planar sheet. These factors make the SiC graphyne and $Si_1C_9$ graphyne be a plane. On the other hand, the $Si_2C_8$ graphyne is a buckled structure with the buckling of 0.05Å, indicating a co-existence of the $sp$, $sp^2$ and $sp^3$ hybridizations. The most interesting finding is that the $sp$ hybridization not only exists between C atoms, but also exists between Si and C atoms, indicating the flexibility of Si-C bonding nature in the SiC systems. Another significant finding is that the electronic structures of the $Si_mC_n$ graphyne are sensitive to the location of the Si atoms. When the six bone positions are occupied by the Si atoms, i.e., the $Si_2C_8$ graphyne, it possesses a gapless material with the Dirac point at the M point, while, if some carbon atoms occupied on the bone position, e.g., the SiC and $Si_1C_9$ graphynes, a band gap opens and the systems behave a semiconductor with the energy gap of 0.955 eV for the SiC graphyne and 0.689 eV for the $Si_1C_9$ graphyne, respectively. The study on thermal stability reveals that the obtained structures are stable below 700 K for the SiC graphyne, 800 K for the $Si_1C_9$ graphyne, and 1000 K for the $Si_2C_8$ graphyne, respectively. They will be destroyed only when the temperature reaches more than 2350 K, 2600 K and 3300 K for SiC graphyne, $Si_1C_9$ graphyne and $Si_2C_8$ graphyne respectively. When the temperature is below these values, the structures can return to the stable structure at 0 K after cooling. These reveals that three new types silicon-carbon graphyne have good thermal stability.

**Acknowledgments**

We would like to acknowledge the funding support for this work received from NSFC (grant No.: 61176118).